\documentclass[11pt]{article}

\usepackage{times}
\usepackage{graphicx}

\title{Strategies for processing diffraction data from randomly oriented particles}
\author{Veit Elser \\ Department of Physics, Cornell University \\ Ithaca, NY 14853-2501 \\ USA}

\begin{document}

\maketitle

\begin{center}
\parbox{4in}{This note compares the single-shot and intensity cross-correlation proposals for x-ray imaging of randomly oriented particles and shows very directly that the latter will usually not be feasible even when the former is. 
} 
\end{center}

\section{Introduction}

The health and growth of any field of science can often be traced to a few, singularly creative and visionary individuals whose contributions are pervasive. Such is the case in the relatively new field of coherent imaging, as the application of even an unsophisticated ``common-element" algorithm to author lists will confirm [1-16]. Scientists in this class succeed by not being overly inhibited by the relentless filters of practicality, plausibility, and even feasibility. This task falls to a lower rank of inquiry, and it is this function that the present contribution is meant to serve.

This note considers the problem of imaging identical, randomly oriented particles with diffraction signals. We first review two imaging schemes that would appear to be quite different in their experimental requirements. This is shown to be false: the feasibility of both methods depends critically on the mean number of photons scattered per particle and one of the schemes is superior in this respect.

Our discussion only considers the reconstruction of the 3D intensity distribution of the particle. The easier problem of reconstructing the particle contrast from the intensity, \textit{i.e.} reconstructing the phase, is treated at length elsewhere \cite{Millane, ElserMillane}.

\section{Two data collection schemes}

X-ray free-electron lasers offer the possibility of imaging individual particles that currently have to be crystallized in order to produce sufficient signal \cite{Neutze}. The increase in brightness over synchrotron sources is so enormous that the best way of utilizing this new resource is still wide open. We will discuss two methods that have been proposed. Common to both methods is a train of short (10 fs) pulses, each comprising a fixed number of photons ($10^{12}$). The methods differ in how the pulses are applied to the particles --- of which we have an unlimited supply --- and how the data is collected.

\subsection{Single-shot data}

In the \textit{single-shot} scheme \cite{Neutze, Loh} the pulses are focussed to a small area $A_{ss}$, such that each pulse hits at most one particle. On average, each hit elastically scatters $N_{ss}$ photons into the range of angles useful for structure determination. The data collected in this scheme are sets of photon counts $K_p(\mathbf{q}_i)$, where $p$ identifies the hit (pulse) and $\mathbf{q}_i$ is the spatial frequency associated with photons scattered into detector pixel $i$. The time between pulses (10 ms) is assumed to be sufficient for detector read-out and initialization for the next pulse. Although the particles are identical in structure (at the resolution of interest), the particle orientation in each hit is random and unknown. The duration of the pulses is short enough that rotational motion and particle explosion (due to inelastic processes) have a negligible effect on the elastically scattered photons.

The goal of the single-shot scheme is to determine the unique 3D intensity distribution $I(\mathbf{q})$ of a single particle that has the highest probability of ``explaining"  all of the data $K_p(\mathbf{q}_i)$ when the particle in each hit is assigned a suitable rotation $\mathbf{R}_p$ and the corresponding photon fluence $\phi_p$ a particular value. To determine $I(\mathbf{q})$ one maximizes the log-likelihood function \cite{LohElser}
\begin{equation}
\sum_p \log{P\left(\{K_p(\mathbf{q}_i)\}_{i=1,\ldots}\;|\; I(\mathbf{q}),\, \mathbf{R}_p,\, \phi_p\right)}
\end{equation}
where $P(\ldots | \ldots)$ is the conditional probability, based on Poisson statistics, of the data from pulse $p$ given the model $I(\mathbf{q})$ and parameters $\mathbf{R}_p$, $\phi_p$.
This intensity reconstruction problem is already interesting when we do not limit the number of hits. In this case it is only the characteristics of the incident pulse, or equivalently the average number of scattered photons $N_{ss}$, that can effect the feasibility of the method. We provide the same resources --- unlimited supply of particles and x-ray pulses --- to the competing scheme (next section).

\subsection{Cross-correlation data}

The alternative scheme is based on intensity \textit{cross-correlations} \cite{Kam, Saldin1}. Here the same pulses are focussed to a much larger area $A_{cc}$, where they scatter from an ensemble of many particles\footnote{As originally proposed \cite{Kam} the particles would be in solution; in free-electron laser experiments the particle ensemble would be prepared in a vacuum.} which we again assume are identical except for orientation which is random and unknown. The average number of photons scattered per particle, relative to the single-shot method, is reduced by the ratio of areas: $N_{cc}=(A_{ss}/A_{cc})N_{ss}$. We assume the focus area $A_{cc}$ is large relative to the transverse coherence length of the pulse so that the diffraction signals of the particles combine incoherently. The raw data in this method are the photon counts $K_p(\mathbf{q}_i)$ and products $K_p(\mathbf{q}_i) K_p(\mathbf{q}_j)$, recorded by the detector after every pulse $p$ scatters from the particle ensemble. Unlike the single-shot scheme, which records data for every pulse (hit), the raw data here are averaged over pulses to obtain the cross-correlations
\begin{equation}\label{ccdata}
C(\mathbf{q}_i, \mathbf{q}_j)=\langle K_p(\mathbf{q}_i) K_p(\mathbf{q}_j)\rangle_p-\langle K_p(\mathbf{q}_i)\rangle_p\langle K_p(\mathbf{q}_j)\rangle_p.
\end{equation}
As in the single-shot scheme we assume the number of pulses and number of particles (in the ensemble) is unlimited so that the errors in these averages can be suitably small. Thus the feasibility of the method can only depend on the average number of photons scattered per particle, $N_{cc}$.

The goal of the cross-correlation scheme is the same as for the single-shot scheme: to determine the unique 3D intensity $I(\mathbf{q})$ of a single particle that is consistent with the data, in this case the cross-correlations. This translates to the set of equations
\begin{equation}\label{cqq'}
C(\mathbf{q}, \mathbf{q'})=\langle I(\mathbf{R}\cdot \mathbf{q}) I(\mathbf{R}\cdot \mathbf{q'})\rangle_\mathbf{R}-\langle I(\mathbf{R}\cdot \mathbf{q})\rangle_\mathbf{R}\langle I(\mathbf{R}\cdot \mathbf{q'})\rangle_\mathbf{R},
\end{equation}
for the function $I(\mathbf{q})$ at all pairs of measured spatial frequencies $\mathbf{q}$ and $\mathbf{q'}$. The averages in this equation are with respect to the rotations $\mathbf{R}$ of the particle axes relative to the incident beam. In the absence of any particle alignment mechanism the distribution of $\mathbf{R}$ is uniform.

The cross-correlation scheme has several practical advantages over the single-shot scheme. Since the diffraction signal recorded from each pulse is proportional to the number of particles in the ensemble, the signal-to-noise in the raw data will be much higher. Not having to focus the beam to a small area and being able to simply average the data over pulses (rather than storing each one) also greatly simplifies the experiment. But these practical considerations are irrelevant in the event that the intensity $I(\mathbf{q})$ cannot be reconstructed at all. The theoretical feasibility of the reconstruction should be the foremost criterion when evaluating the two schemes.

\section{Comparison of the two schemes}

As a first step in comparing the two schemes we show that the single-shot data could be processed in the manner of the cross-correlation data but not conversely. Moreover, since the number of photons scattered per particle is greater in the single-shot scheme, in the absence of practical factors this is the superior method for collecting data.

Because the photons scattered by the particle ensemble combine incoherently, we can express the counts recorded by the detector from pulse $p$ in the form
\begin{equation}
K_p(\mathbf{q}_i)=\sum_\alpha K_{p\alpha}(\mathbf{q}_i),
\end{equation}
where the index $\alpha$ identifies particles in the ensemble. If there were an oracle\footnote{Oracles are often used in theoretical computer science to isolate particular parts of a problem. It is in this sense that the oracle is used in present context.} that somehow was able to assign particle-origin identifiers $\alpha$ to all the detected photons, then the added information would transform the data in a cross-correlation experiment into data that would be collected in a single-shot experiment, although with a large reduction in the number of scattered photons per particle ($N_{cc}$ vs. $N_{ss}$). Such an oracle, however, does not exist, and so the cross-correlation scheme is always subject to a large information deficit relative to the single-shot scheme.

We can develop the relationship between the two schemes further. Since the orientations of two particles in the ensemble are uncorrelated, the pulse average for distinct particles $\alpha\ne\beta$ vanishes:
\begin{equation}
0=\langle K_{p\alpha}(\mathbf{q}_i) K_{p\beta}(\mathbf{q}_j)\rangle_p-\langle K_{p\alpha}(\mathbf{q}_i)\rangle_p\langle K_{p\beta}(\mathbf{q}_j)\rangle_p
\end{equation}
The cross-correlations (\ref{ccdata}) therefore take the form
\begin{equation}\label{reducedC}
C(\mathbf{q}_i, \mathbf{q}_j)=\sum_\alpha\left( \langle K_{p\alpha}(\mathbf{q}_i) K_{p\alpha}(\mathbf{q}_j)\rangle_p-\langle K_{p\alpha}(\mathbf{q}_i)\rangle_p\langle K_{p\alpha}(\mathbf{q}_j)\rangle_p\right).
\end{equation}
This equation shows that the data processed by the cross-correlation scheme can be thought of as arising from single particle hits, where the number of pulses is simply multiplied by the number of particles in the ensemble (sum over $\alpha$). From the viewpoint of the data that gets processed, the cross-correlation experiments thus look like single-shot experiments. What really distinguishes the two methods is (\textit{i}) how the ``single-shot" data is processed, and (\textit{ii}) the difference in the number of scattered photons per particle ($N_{ss}$ vs. $N_{cc}$).

The cross-correlation scheme has a disadvantage relative to the single-shot scheme both because of restrictions in the processing and the smaller number of photons scattered per particle. These shortcomings are not unrelated. To illustrate this point we suppose that we have a particle that scatters on average $N_{ss}=1000$ photons in the single-shot scheme, and that the linear size of the beam focus in the cross-correlation scheme is larger by a factor of 100; thus $N_{cc}=(1/100)^2 N_{ss}=0.1$. By Poisson statistics 0.5\% of the particles in the cross-correlation experiment will scatter two or more photons, the minimum required to contribute to the cross-correlation (\ref{reducedC}). To escape the limitations imposed by quadratic cross-correlations one could also form cubic correlations 
\begin{equation}
C(\mathbf{q}_i,\mathbf{q}_j,\mathbf{q}_k)=\langle K_p(\mathbf{q}_i) K_p(\mathbf{q}_j)K_p(\mathbf{q}_k)\rangle_p-\cdots
\end{equation}
and higher. But for these correlations only $0.015\%$ of the particles contribute (three or more photons) and higher orders decay very rapidly. The small value of $N_{cc}$ thus limits the experiment to low order correlations.

The question of how many scattered photons (per particle) are required to reconstruct the 3D intensity of a particle, from randomly oriented diffraction data, is an interesting one \cite{Elsercrypto}. There are really two questions: how many photons are required to provide the necessary information, irrespective of the computational complexity of the reconstruction; and, how many photons are needed by a practical reconstruction algorithm? For the second question we already have some good answers. First, we note that single-shot data \textit{could} be processed in the style of a cross-correlation experiment, by simply averaging products of photon counts and forming quadratic, cubic, etc. cross-correlations. However, this data 
reduction/compression step is not done in the most successful, likelihood maximization, algorithms \cite{Fung, LohElser}. By contrast, the latter type of algorithm utilizes all the correlations in the data in the process of iteratively refining a model intensity based on comparisons with all the diffraction patterns. The difficulty of these reconstructions, as measured in the number of iterations, grows dramatically when the number of detected photons per particle drops below a certain value \cite{LohElser}. This number depends weakly on resolution, and for a particle that measures 16 voxels in diameter is about 50 (see Figure 1). These observations combined make us strongly doubt the potential for determining the 3D intensity from noisy, randomly oriented diffraction data by the cross-correlation method.

\begin{figure}[t]
\begin{center}\scalebox{1.2}{\includegraphics{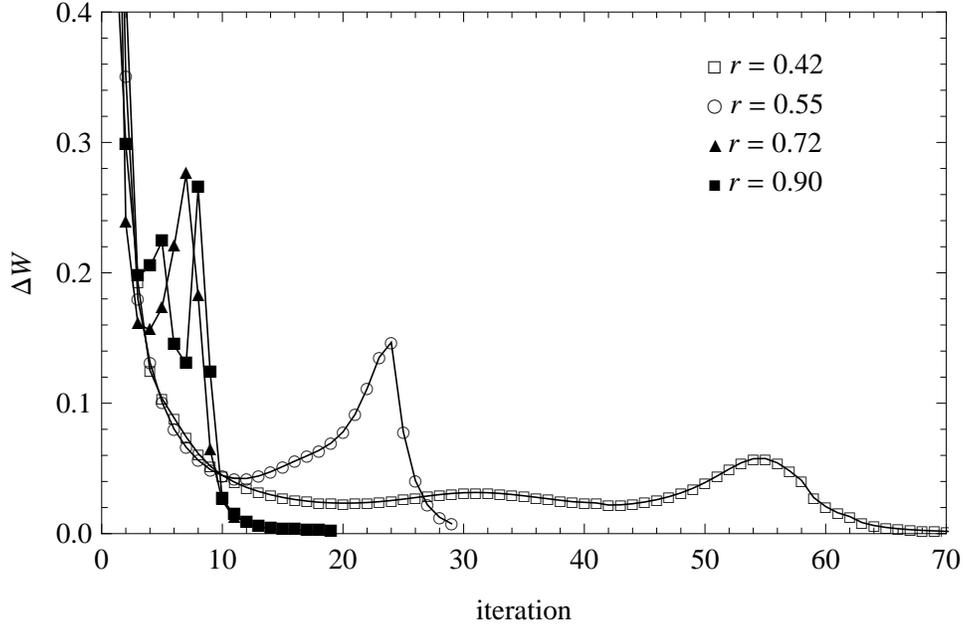}}
\end{center}
\begin{center}
\parbox{6in}{\caption{Update magnitude vs. iteration in a likelihood maximization reconstruction (EMC algorithm \cite{LohElser}) of the intensity of a particle. The four plots correspond to different average numbers of photons $N$ in the diffraction data, and are labeled by their reduced information rates \cite{Elsercrypto}, $r(N)$. The latter (a number between zero and one) is the ratio of the information rate of diffraction data in an experiment with unknown particle orientations to the rate when particle orientations are known. The value $r=1/2$ corresponds to the case where the information in each data with unknown orientation equals the information it provides about the particle orientation. The selection of photon numbers shown, $r(25)=0.42, r(45)=0.55, r(80)=0.72$ and $r(225)=0.90$ emphasizes the sudden onset of slow convergence when the reduced information rate falls below $1/2$. The corresponding photon number, in this case $N\approx 50$, is weakly resolution dependent. The data for these reconstructions were derived from a particle measuring 16 voxels in its diameter.}}
\end{center}
\end{figure}

For the more academic question, regarding information sufficiency of the cross-correlation data, we have a partial answer. Consider a 3D intensity that measures $D$ speckles in diameter. The reconstruction of this intensity corresponds to the discovery of a unique set of real parameters whose number scales as $D^3$. We can evaluate the feasibility of this task by finding the scaling of the number of real-valued constraints provided by the cross-correlations. Because of rotational averaging, the quadratic cross-correlation function reduces to a function of only three real-number arguments:
\begin{equation}
C(\mathbf{q}, \mathbf{q'})=\widetilde{C}(\theta,\theta',\phi).
\end{equation}
Here $\theta$, $\theta'$ and $\phi$ locate $\mathbf{q}$ and $\mathbf{q'}$ on the Ewald sphere in rotation-invariant terms, i.e. two scattering angles and an azimuthal angle about the beam axis\footnote{Symmetry considerations apply to the range of the arguments of the reduced function. A detailed analysis can be found in the appendix.}. This shows that the number of constraints also scales as $D^3$. An answer to the feasibility question thus requires a more detailed analysis. A similar situation arises in connection with the standard phase problem \cite{ElserMillane}, \textit{i.e.} reconstructing particle contrast from autocorrelation (Fourier intensity) data. There the number of free variables and constraints also scale with the same power of resolution and uniqueness depends on the number of dimensions and the shape of the particle support. In the appendix we show that, for the case of a spherical particle, the number of free variables in the contrast outnumber by a factor of 1.43 the independent constraints provided by the quadratic cross-correlations\footnote{This conclusion is stronger than that obtained by Kam \cite{Kam} and Saldin \textit{et al.} \cite{Saldin1}.}. A similar counting argument, when applied to an ensemble of particles aligned along the incident beam axis, shows that quadratic cross-correlations have sufficient information to reconstruct a 2D intensity when only one particle angle is unknown\footnote{The scaling of the number of speckles in the 2D intensity, $D^2$, is exceeded by the scaling of the number of cross-correlations, $D^3$.}. This has been confirmed in simulations \cite{S16}. On the other hand, information sufficiency for the unaligned case could be achieved with the cubic cross-correlations. Even with rotational averaging taken into account, the number of independent cubic correlations scales as $D^5\;$\footnote{The reduced function depends on three scattering angles and two azimuthal angles about the incident beam.}. Since photon numbers as low as $N_{cc}=1$ are adequate for obtaining good cubic cross-correlations, while such low numbers are out of reach of likelihood maximization algorithms, it appears that intensity reconstruction from cubic cross-correlations is a very difficult problem.

\section{Conclusions}

Whereas the cross-correlation scheme for collecting structure data from randomly oriented particles has practical advantages over the single-shot scheme, these advantages do not compensate for the information deficit that distinguishes the two schemes. Cross-correlation experiments are an interesting and appropriate use of the high photon flux in X-ray free-electron laser sources. However, their use in elucidating structure will likely be limited to statistical properties \cite{Treacy, Wochner} and fall short of complete structure recovery.

\section{Appendix: data insufficiency for spherical particles}

Consider a particle with real density $\rho(\mathbf{r})$. The support of $\rho(\mathbf{r})$ is spherical with radius $R$ in units where the sampling density, set by the resolution, is 1. The number of density variables to be reconstructed from our data is therefore $V=(4\pi/3) R^3$.

We can avoid sampling issues associated with the particle intensity $I(\mathbf{q})$ --- a bandlimited function --- by working instead with its Fourier transform: the density autocorrelation:
\begin{equation}\label{auto}
A(\mathbf{r})=\int d^3\mathbf{r'}\rho(\mathbf{r'})\rho(\mathbf{r'}+\mathbf{r}).
\end{equation}
The autocorrelation is a real function with the symmetry property
\begin{equation}\label{inversionsym}
A(\mathbf{r})=A(-\mathbf{r}).
\end{equation}
Since $A(\mathbf{r})$ has a spherical support of radius $\tilde{R}=2R$ we have
\begin{equation}\label{samplenumber}
\frac{1}{2}\frac{4\pi}{3}\tilde{R}^3
\end{equation}
independent autocorrelation samples.

To work with the rotational averages in the data we switch to an angular momentum basis for $A(\mathbf{r})$. Writing $\mathbf{r}=r\mathbf{n}$ in terms of its magnitude $r$ and direction $\mathbf{n}$, we have
\begin{equation}\label{angbasis}
A(\mathbf{r}) = \sum_{l=0, 2, 4,\ldots}\;\sum_{m=-l}^l A_{l m}(r) Y_{lm}(\mathbf{n})
\end{equation}
where the restriction to even $l$ comes from (\ref{inversionsym}).

Taking the Fourier transform of (\ref{cqq'}) with respect to each argument we obtain
\begin{equation}\label{tildeC}
\widetilde{C}(\mathbf{r},\mathbf{r'})=\langle A(\mathbf{R}\cdot \mathbf{r}) A(\mathbf{R}\cdot \mathbf{r'})\rangle_\mathbf{R}-\langle A(\mathbf{R}\cdot \mathbf{r})\rangle_\mathbf{R}\langle A(\mathbf{R}\cdot \mathbf{r'})\rangle_\mathbf{R},
\end{equation}
where the averages are with respect to rotations $\mathbf{R}$ of the particle. Expressing $A(\mathbf{R}\cdot \mathbf{r})$ in terms of the angular momentum basis introduces the replacement
\begin{equation}
Y_{lm}(\mathbf{n})\to \sum_{m''=-l}^l D_{m m''}^l(\mathbf{R})Y_{l m''}(\mathbf{n})
\end{equation}
in (\ref{angbasis}), where $D(\mathbf{R})$ is the rotation matrix for $\mathbf{R}$ in the angular momentum basis. Using the fact that $A(\mathbf{r})$ is real and the orthogonality property (averaging in  the uniform distribution of $\mathbf{R}$)
\begin{equation}
\langle D_{m m''}^l(\mathbf{R}) D_{m' m'''}^{l'}(\mathbf{R})^\ast\rangle_\mathbf{R}=\frac{1}{2l+1}\delta_{l l'}\delta_{m m'}\delta_{m'' m'''},
\end{equation}
equation (\ref{tildeC}) takes the form
\begin{equation}
\widetilde{C}(\mathbf{r},\mathbf{r'})=\sum_{l=2, 4, 6,\ldots}\frac{1}{2l+1}\sum_{m=-l}^l A_{lm}(r) A_{lm}^\ast(r') \sum_{m'=-l}^l Y_{lm'}(\mathbf{n}) Y_{lm'}^\ast(\mathbf{n'}).
\end{equation}
Applying the addition theorem on the sum over $m'$ we obtain the result
\begin{equation}\label{Crr'theta}
\widetilde{C}(r,r',\theta)=\frac{1}{4\pi}\sum_{l=2, 4, 6,\ldots} P_l(\cos{\theta}) \sum_{m=-l}^l A_{lm}(r) A_{lm}^\ast(r'),
\end{equation}
where $\cos{\theta}=\mathbf{n}\cdot \mathbf{n'}$ and $P_l$ is the Legendre polynomial. 

The next task is to count the number of independent cross-correlation constraints (\ref{Crr'theta}) given that the autocorrelation $A(\mathbf{r})$ is sampled at a finite resolution. Our approach is to express $A(\mathbf{r})$ in the basis of solutions to the mode equations
\begin{equation}
-\nabla^2 A_{k l m}(\mathbf{r})=q_{k l m}^2 A_{k l m}(\mathbf{r})\qquad A_{k l m}(\mathbf{r})=0,\,  r>\tilde{R}
\end{equation}
and thereby arrive at a discrete set of equations. As above, $l$ and $m$ are angular momentum quantum numbers; the new index $k=1,2,\ldots$ identifies the radial structure of the mode. By including all modes with the property
\begin{equation}\label{qbound}
q_{k l m}<Q
\end{equation}
for an appropriate $Q$, our basis for $A(\mathbf{r})$ has a finite, isotropically defined resolution. Applying standard mode counting to modes of bounded linear momentum $|\mathbf{q}|<Q$ within a spherical box of volume $\tilde{V}=(4\pi/3)\tilde{R}^3$,
\begin{equation}
\int_0^Q\frac{\tilde{V}}{(2\pi)^3}\,4\pi q^2 dq=\tilde{V},
\end{equation}
we obtain
\begin{equation}
Q=(6\pi^2)^{1/3}.
\end{equation}
The solutions to the mode equations have the form (\ref{angbasis}), where the radial functions are expanded in terms of spherical Bessel functions:
\begin{equation}\label{klm}
A_{l m}(r)=\sum_{k=1}^{k(l)}A_{k l m}\, j_l(q_{k l}\,r).
\end{equation}
We have suppressed the angular momentum index $m$ on the mode eigenvalues $q_{k l m}$ since the latter are determined by the $m$-independent radial equation:
\begin{equation}
\left[-\frac{1}{r}\left(\frac{d}{dr}\right)^2 r+\frac{l(l+1)}{r^2}\right]j_l(q_{k l}\,r)=q_{k l}^2\, j_l(q_{k l}\,r)\qquad j_l(q_{k l}\tilde{R}) = 0.
\end{equation}
The maximum radial index $k(l)$ is defined as the maximum $k$ that satisfies (\ref{qbound}).

For the purpose of counting constraints the only thing we need to know about the modes $A_{k l m}(\mathbf{r})$ is the number of radial modes $k(l)$ for each angular momentum index $l$. In the limit of many modes ($\tilde{R}\to\infty$) we estimate $k(l)$ using the WKB approximation for the spherical Bessel function,
\begin{equation}\label{WKB}
j_l(q_{k l}\,r)\sim B_l\, \frac{\sin{\phi_{k l}(r)}}{r},
\end{equation}
valid for $r>r_{k l}$ where $r_{k l}= \sqrt{l(l+1)}/q_{k l}\sim l/q_{k l}$ is the turning point. For fixed $l$, the highest radial mode $k(l)$ is the one with the smallest turning point:
\begin{equation}
r_{k l}\ge r_{k(l) l}=\frac{l}{Q}= r(l).
\end{equation}
Up to constant turning point corrections, the phase $\phi_{k l}(r)$ of the sine function in (\ref{WKB}) runs through $k(l)$ multiples of $\pi$ between this smallest turning point and $r=\tilde{R}$:
\begin{equation}
\pi k(l)\sim \phi_{k(l) l}(\tilde{R})-\phi_{k(l) l}(r(l)) =\int_{r(l)}^{\tilde{R}}\sqrt{Q^2-\frac{l^2}{r^2}}\, dr=\tilde{R}Q\,f(l/\tilde{R}Q)
\end{equation}
\begin{equation}
f(x)=\sqrt{1-x^2}-x\, \arccos{x}.
\end{equation}
The combination
$L=\tilde{R}Q$
is identified with the largest possible $l$, where there is just a single mode with smallest turning point $r(L)$ equal to $\tilde{R}$.

Taking advantage of mode orthogonality in (\ref{Crr'theta}) we define
\begin{equation}\label{Ckk'l}
\widetilde{C}(k,k',l)=\int_0^{\tilde{R}}j_l(q_{k l}\,r)\, r^2 dr \int_0^{\tilde{R}}j_l(q_{k' l}\,r')\, r'^2 dr' \int_0^\pi P_l(\cos{\theta})\, 2\pi\sin{\theta}\, d\theta\;\; \widetilde{C}(r,r',\theta)
\end{equation}
and using (\ref{klm}) obtain the discrete set of constraint equations for the mode amplitudes of the density autocorrelation,
\begin{equation}\label{constraints}
\widetilde{C}(k,k',l)=N_{k k' l}\sum_{m=-l}^l A_{klm} A_{k'lm}^\ast,
\end{equation}
where the $N_{k k' l}$ are normalization constants:
\begin{equation}
N_{k k' l}=\frac{1}{2l+1}\left(\int_0^{\tilde{R}}j_l^2(q_{k l}\,r) \,r^2 dr\right)\left(\int_0^{\tilde{R}}j_l^2(q_{k' l}\,r') \,r'^2 dr'\right).
\end{equation}

Both Kam \cite{Kam} and Saldin \textit{et al.} \cite{Saldin1} obtain equations of the form (\ref{constraints}) but where the spherical harmonic expansion of the density autocorrelation $A(\mathbf{r})$ is replaced by the expansion of the intensity $I(\mathbf{q})$. These authors also remark that since the constraint equations for different $l$ are completely decoupled, the reconstruction is ambiguous up to arbitrary relative rotations applied to the different principal angular momentum ($l$) components of the intensity (or equivalently, the density autocorrelation). This conclusion, however, is based on the premise that the autocorrelation samples are the independent variables of the reconstruction problem when in fact the true independent variables are the density samples. In the case of our spherical particle, for example, the number of autocorrelation samples exceeds the number of density samples by a factor of four. When the constraint equations (\ref{constraints}) are expressed in terms of the spherical harmonic expansion of $\rho(\mathbf{r})$ using (\ref{auto}), the different principal angular momentum components of $\rho$ are coupled in the resulting quartic equations.

The main advantage in expressing the constraint equations (\ref{constraints}) in terms of the autocorrelation, rather than the intensity, is that it is possible to count the actual number of constraints. Clearly the data (\ref{Ckk'l}) are symmetric with respect to interchanging $k$ and $k'$. We also know that these indices range between $1$ and $k(l)$ for $l$ values up to $L=\tilde{R}Q$ and that the symmetry (\ref{inversionsym}) restricts us to only even $l$. Altogether the number of independent constraint equations is therefore given by:
\begin{equation}
E=\sum_{l=2, 4, 6,\ldots}^{L}\;\;\sum_{k=1}^{k(l)}\;\;\sum_{k'=k}^{k(l)}1\sim\frac{L^3}{4\pi^2}\int_0^1f^2(x)\, dx=\left(\frac{8}{9}-\frac{\pi}{6}\right)\tilde{R}^3.
\end{equation}
Comparing with the number of free variables, $V=(\pi/6)\tilde{R}^3$, we see that the reconstruction problem is underconstrained:
\begin{equation}
E/V=\frac{16}{3\pi}-1\approx 0.698.
\end{equation}

\end{document}